\documentclass[a4paper,twoside]{article}
\usepackage[authoryear]{natbib}
\usepackage{epsfig}
\usepackage{graphicx}
\usepackage{color}
\bibpunct{(}{)}{;}{a}{}{,}
\date{} %Please leave the date blank
\newcommand{\kms}{\mbox{km\,s$^{-1}$}}
\newcommand{\affil}[1]{$^{\rm #1}$}

\newcommand{\arcsec}{\ensuremath{\rm arcsec}}%{^{\prime\prime}\!\!\!#1\,}
\newcommand{\kpc}{\ensuremath{\rm kpc}}
\renewcommand{\mag}{\ensuremath{\rm mag}}

\newcommand{\objsev}{\rm Cyg~OB2~\#7}%{\rm Cyg~OB2~\textnumero7~}
\newcommand{\objeleven}{\rm Cyg~OB2~\#11}%{\rm Cyg~OB2~\textnumero11~}
\newcommand{\objelevenn}{\rm \bf Cyg~OB2~\#11}
\newcommand{\Msun}{\ensuremath{\rm M_\odot}}

\newcommand{\href}[1]{\url{#1}}
\newcommand{\apjjj}{ApJ}%{Astrophys. J.} %  Astrophysical Journal
\newcommand{\ab}{Astrophysical Bulletin}
\newcommand{\aaa}{A\&A}%{Astronom. and Astrophys.}
\newcommand{\apj}{Astronom. J.}
\newcommand{\apjl}{Astrophys. J. Letters}
\newcommand{\apjs}{ApJS}%{Astrophys. J. Suppl.}
\newcommand{\mnras}{MNRAS}%{Monthly Notices Roy. Astronom. Soc.}
\title{\large\bf\flushleft Investigation of  \objelevenn \ \bf (O5 Ifc) by modeling its atmosphere}
\author{\parbox{\textwidth}{\flushleft
\vspace{-0.5cm}
{\it Olga Maryeva\affil{A,C}, Roman Zhuchkov\affil{B} and Eugene Malogolovets\affil{A}}\\
\vspace{0.4cm}
{\small \affil{A}\,Special Astrophysical Observatory, Russian Academy of Sciences, Nizhnii Arkhyz 369167, Russia}\\
{\small \affil{B}\,Astronomy and  geodesy department of Kazan (Volga region) Federal University,\\ Kremlevskaya str., 18, Kazan, 420008, Russia}\\
{\small \affil{C}\,Email: olga.maryeva@gmail.com}}}
\begin{document}
\twocolumn[
\begin{changemargin}{.8cm}{.5cm}
\begin{minipage}{.9\textwidth}
\vspace{-1cm}
\maketitle
%%%%%%%%%%%%%     ABSTRACT    %%%%%%%%%%%%%
\small{\bf Abstract:}
      We continue the study of O-supergiants belonging to the association
      Cyg~OB2 using moderate-resolution spectra.
      In this paper we present results of the modeling of the stellar
      atmosphere of \objeleven.
%может добавить -- по спектрам в ультрафиолетовой и оптической области????
%%%%
      This object belongs to the spectral class ${\rm Ofc}$, which was
      recently introduced and yet small in numbers. ${\rm Ofc}$ class consists
      of stars with normal spectra
      with C{\scriptsize III}~$\lambda\lambda4647,4650,4652$ emission lines of comparable intensity to those of
      the ${\rm Of}$-defining lines N{\scriptsize III}~$\lambda\lambda4634,4640,4642$.
      We combined new spectral data obtained by the 1.5-m Russian-Turkish telescope with spectra from MAST and CASU archives
      and   determined physical parameters of the wind and chemical composition of the stellar atmosphere using {\sc cmfgen} code.
    %  Chemical composition of \objeleven \ differs  from the one of normal O4-6 supergiant,
%%%%
     The estimated nitrogen abundance is lower  than one in atmospheres of ``normal'' O-supergiants (i.e. O4-6 supergiants without additional spectral index ``n'' or ``c'') and carbon abundance  is solar.
%\footnote{\textcolor{green}{Although the word ``normal'' is not quite applicable to high luminosity stars, in this article ``normal'' will mean O4-6 supergiants without additional spectral index ``n'' or ``c''. }} and carbon abundance  is solar.} % almost solar.
%      В атмосфере наблюдается более низкое по сравнению с данными звёздами содержание азота, и почти солнечное содержание углерода.
      Also we find an excess in silicon. %%PA
We present an illustrative comparison of our modeling results
     with current Geneva evolutionary models for rotating massive stars.
     The position  on the Hertzsprung-Russell diagram corresponds to the star  mass
      of about 50~\Msun \ and age about %%PA
     4.5 Myr. Moreover, we carried out the high angular resolution ($\sim0.02$
     \ \arcsec) observations on the Russian 6-meter %%PA
     telescope aiming to find weaker companions of this star, which did not
     reveal any.

%Так же мы обнаружили значительный избыток кремния.

%%% Что ещё добавить??? может какие-то слова -- что наша работа добавляет данные для статистики???

%%%%      found anomalies in chemical composition.
%%%%      Chemical composition of \objeleven \ differs significantly from the one of   \objsev \ supergiant studied previously.
%%%%      %The chemical composition of \objeleven \ differs significantly from previously studied chemical
%%%%      %composition of the supergiant \objsev.
%%%%      There is a considerable excess of carbon and silicon
%%%%      in the atmosphere of \objeleven, while in atmosphere of \objsev \ abundances of carbon and silicon are solar.

%?????????????????????.
%%%%%%%%%%%%%     KEYWORDS    %%%%%%%%%%%%%

%%%%%%%%DO NOT EDIT%%%%%%%%%%%%
\medskip
\medskip
\end{minipage}
\end{changemargin}
]
\small
%%%%%%%%EDIT FROM HERE%%%%%%%%%%%%
\section{Introduction}\label{introduction}

       Improvement in telescope detection limits, %%PA
 technical evolution of detectors and emergence of
       huge databases of observational data  resulted in more careful spectral classification of objects,
       and provision of new spectral subclasses. %%PA provision --> introduction?
In 2010, \citet{WalbornOIfc} proposed to introduce a
       new subclass ${\rm Ofc}$ to denote O-stars with comparable intensity of C{\scriptsize III}~$\lambda\lambda 4647,  4650, 4652$
       and N{\scriptsize III}~$\lambda\lambda4634, 4640, 4642$ lines. This phenomenon is often observed at
       spectral type O5 at all luminosity classes, but preferentially in some
       associations or clusters and not the others \citep{WalbornOIfc}. %%PA
       For today, eighteen Galactic O-stars are classified as ${\rm Ofc}$. %%PA

Although  ${\rm Ofc}$ class was introduced recently, CNO dichotomy is well known among O-stars (\citet{WalbornCNO,Walborn03,Walborn76} and references therein). Anticorrelations of N
versus C and O and correlations with He/H %in the expected sense
have encouraged interpretations in terms of mixing of
CNO-cycled material into the atmospheres and winds of massive stars. 
The mixing depends on the rotation rate and increases at low metallicity \citep{MaederMeynet}.
But whether ${\rm Ofc}$ stars are related to CNO dichotomy only? With the advent of the ${\rm Ofc}$ class of stars the following questions arose:
\begin{itemize}
%%%      \item[-] Why does the excess  of carbon appear?
      \item[-] Whether it is related only to the CNO-cycle and mixing processes in the
               star itself or it arises under the influence of a general dynamical evolution
               of clusters or associations?
      \item[-] How does multiplicity of objects influence the excess of carbon?
      \item[-] Are there differences between the physical parameters of ${\rm
          Ofc}$ stars and ${\rm Of}$ stars?
      \item[-] How should an $ {\rm Ofc} $ star evolve further? %%PA
\end{itemize}
%%%%%%%%%%%%%%%%%%%%%%%%%%%%%%%%%%%%%%%%%%%%%%%%

To obtain more data and to estimate more parameters of ${\rm Ofc}$ stars are
important steps for better understanding of the nature of these objects.
%Numerical modeling of atmospheres may employ for this
%Может здесь сказать не о численном моделировании, а о наборе статистики?
%Что изучение таких объектов, расположенных в различных областях, определение их физических параметров, химического состава, а также эволюционного статуса -- важно для понимая природы всего класса этих объектов.
%?????????????????????????????
%%%       Numerical modeling of atmospheres of ${\rm Ofc}$ stars may help answering these questions.

%%% Это статья посвящена сверхгиганту и основная цель этой статьи -- добавить данных по $ {\rm Ofc}$ звёздам.

       \objeleven\footnote{$\alpha$=20:34:08.52 $\delta$=+41:36:59.36 according to SIMBAD {http://simbad.u-strasbg.fr/simbad/}} \ is one of the Galactic supergiants %classified as $ {\rm O5.5~Ifc}$ \citep{WalbornOIfc,Sota}.        This star is
       located  near the northern border of the Cyg~OB2 (VI Cygni) association. This association was first
       noticed more than half a century ago by \citet{MunchMorgan}. It still attracts the attention of researchers
       due to the large number of O-stars and extremely high, heterogeneous
       interstellar reddening. Star \#11
       was immediately recognized as a member of the association \citep{MunchMorgan}. The first spectroscopy of
       the brightest stars  belonging to Cyg~OB2 (including \#11) was performed by  \citet{JohnsonMorgan}. They
       classified \objeleven \ as O6f. Later \citet{Walborn73} classified the star as ${\rm O5~If_+}$ using spectrograms
       obtained at Kitt Peak National Observatory. \citet{MT91} performed its CCD photometry in three (U, B, V) bands.
% выполнили CCD фотометрию ассциации в трёх полосах.
       \objeleven \ is included in their catalog under number 734 (MT91 734).
       Stellar magnitude in $V$ band is $V=10.03~{\rm mag}$ \citep{MT91}. %     \objeleven входит в их каталог под номером 734 (MT91 734)
       As of now, \objeleven \ is classified as $ {\rm O5.5~Ifc}$ \citep{WalbornOIfc,Sota}. \citet{binaryO11} found that \objeleven \ is a single-lined
       spectroscopic binary (SB1 type).

       The spectrum of \objeleven \ was modeled previously by several groups.
       Initially, by means of numerical modeling \citet{Herrero1999} determined
       the parameters of \objeleven \ (effective temperature, luminosity, ${\log{g}}$
       and helium abundance).  They compared the spectral line profiles of H, He I
       and He II with the line profiles synthesized for a large set of NLTE plane-parallel,
       hydrostatic model atmospheres.  \citet{HerreroUV} measured terminal velocity of
       the wind employing resonance lines in the ultraviolet range. Then \objeleven \
       was modeled by \citet{Herrero2002} using the {\sc fastwind} \citep{fastwind, Puls} code.
       In that work the mass-loss rate and velocity law in the stellar wind have been
       determined. Finally, \citet{MokiemObj7} for the first time applied the automated
       fitting method and clarified the physical parameters of \objeleven. It is worth
       noting that the mass-loss rate was determined by \citet{Herrero2002} and \citet{MokiemObj7}
       without taking clumping into account. Inhomogeneities in the winds of the stars were
       studied in the article \citet{PulsMarkova}.  Based on a simultaneous modeling of
       H$\alpha$, infrared, millimeter and radio observations authors concluded that
       clumping is three to six times stronger in the lower wind, where H$\alpha$ forms,
       compared with the outer wind, where the radio continuum originates.

%%%%%%%%%%%%%%%%%%%%%%%%%%%%%%%%%%%%%%%%%%%%%%%%
       In the next section, we describe the observational data and their processing. In
       Section~\ref{sec:model} we will tell about the construction of the model,
       discuss the results and compare them with previous works.
       Section~\ref{sec:chemical} is devoted to  determination of chemical composition of the atmosphere of \objeleven,
       while Section~\ref{sec:diagram} shows locations of  \objeleven \ on the different diagrams.
       The search for companions of \objeleven \ is described in Section~\ref{sec:binary}.
       The conclusions are presented in Section~\ref{sec:results}.

\section{Observational Data}\label{sec:obs}

       We have combined spectra  of \objeleven \ from the archives of 4.2-m William Herschel
       Telescope (WHT)\footnote{{http://casu.ast.cam.ac.uk/casuadc/ingarch/query} This paper
       makes use of data obtained from the Isaac Newton Group Archive which is maintained as
       part of the CASU Astronomical Data Centre at the Institute of Astronomy, Cambridge.}
       with new data obtained on the 1.5-m Russian-Turkish telescope (RTT150). Archival spectra
       were obtained with the ISIS (the Intermediate dispersion Spectrograph
       and Imaging System) instrument of the WHT %%PA
       in July 1995 and in September 1998. R600B and R1200R gratings were used for observations.
       The spectral ranges are 4000-4800 \AA\AA \ and 6350-6750 \AA\AA, and spectral resolutions
       are $\sim3$~\AA \ and $\sim1.5$~\AA, correspondingly. Detailed description of these data
       can be found in paper by \citep{WalbornHowarth}. We processed the data using standard
       procedures for a long-slit spectroscopy.

       Echelle spectrum of \objeleven \ was obtained in November 2012 with the
       TFOSC (TUBITAK Faint Object Spectrograph and Camera) instrument, installed
       in the Cassegrain focus of RTT150.  The
       spectral resolution is $\lambda/\Delta\lambda=2500$ and spectral range is
       4200-8000 \AA\AA.    {\sc dech}  software package was used for the data reduction and analysis \citep{dech}.
       {\sc dech} package includes all the standard stages of echelle data reduction
       process.  Methods of observations and data reduction are the same as
       described by us in the paper~\citep{me}.  Besides, investigated object was
       observed in February 2013 on the same instrument with comparable integration
       time and image quality.  Total exposure time was 80 minutes in both cases. No
       significant differences were found between the spectra obtained in 2012 and 2013
       at an interval of 80 days. In the overall spectrum the signal-to-noise ratio per sampling element is
       ${\rm S/N}=100$ in the blue part (5000 \AA) and 200 in the red part (7000
       \AA). %%PA: signal-to-noize per resolution or sampling element?

       Moreover we have performed observations on the 6-m telescope of Special Astrophysical
       Observatory (SAO) with the speckle interferometer during December 2012 to
       look for close components at distances $0.02-4 \ \arcsec$. These observations
       are described in more details in Section~\ref{sec:binary}.

%      !!!!!!!!!!!!!!!!!!!!!!!!!!!!!!!!!!!!!!!!!!1
       We investigated the object in the UV range using the spectra obtained by the
       {\it Hubble Space Telescope (HST)} with the STIS spectrograph and published by
       \citet{HerreroUV}. These data were taken from the  Multimission
       Archive at STScI (MAST)\footnote{{http://archive.stsci.edu/}}. The spectral
       range of these data is $1150-1700$ \AA\AA, and the spectral
       resolution is $\lambda/\Delta\lambda \sim 1000-1500$, ${\rm  S/N=19} $.

\section{Modeling}\label{sec:model}

       In our work  atmospheric parameters of \objeleven \ are       %%%   at the first time
       determined using both ultraviolet and optical data simultaneously. %%%   at the first time
       We have used the {\sc cmfgen} atmospheric modeling code
       \citep{Hillier5}. % to determine physical parameters of the atmosphere of       \objeleven.
       This code solves radiative transfer equation for objects with
       spherically symmetric extended outflows using either the Sobolev approximation or
        the full comoving-frame solution of the radiative transfer equation.
       {\sc cmfgen} incorporates line blanketing, the effect of Auger ionization and
       clumping. Every model is defined by a hydrostatic stellar radius $R_*$,
       luminosity $L_*$, mass-loss rate $\dot{M}$, filling factor $f$, wind terminal
       velocity $v_\infty$, stellar mass M and by the abundances $Z_i$ of included
       elementary species.
    Although in {\sc fastwind} ions of different chemical elements are taken into account,
    {\sc fastwind} models computed by \citet{Herrero2002, MokiemObj7} 
  %  assumed  solar abundances for all elements except for H and He
%!!!!!!!!!!!!!!!!!!!!!!!!!!!!!
   assumed a solar mix of metals, and therefore %, He and Si, and therefore
    could not be used to determine the chemical composition of stellar
    atmosphere. %%PA: ??
    As a result, in our work we for the first time determine the chemical abundance of the atmosphere of \objeleven.

    In the previous studies the absolute magnitude (${\rm M_v}$) and  bolometric    corrections (${\rm BC_v}$)
    %, which depends on bolometric    corrections (${\rm BC_v}$),
    were employed to estimate the luminosity.  Numerous tabulations of    ${\rm BC_v}$ and
    ${\rm M_v}$ for OI~5 existing in the literature sometimes differ significantly.
    For example, the absolute magnitude of \objeleven \ in the work of \citet{MT91} is ${\rm
    M_v=-6.9~\mag}$, in articles of \citet{MokiemObj7} and \citet{Herrero1999} it is ${\rm
    M_v=-6.51~\mag}$, while in \citet{PulsMarkova}  ${\rm M_v=-6.67~\mag}$. At the same time,
    the absolute magnitude of O5~I class stars is $ {\rm M_v = -6.33~\mag}$, according to
    calculations of \citet{MartinsBC}, the same value as assumed by \citet{KiminkiAv} for \objeleven.
     We will use other method of determining the luminosity. To accurately determine the luminosity
     of the object, the magnitudes of the star were calculated in the U, B,  V  and R filters
     from the model spectra and compared with observations. In order to obtain the magnitudes
     for the model spectra, we first recomputed the fluxes for the distance to the
     Cyg~OB2 association (1.5 \kpc \ according to \citet{KiminkiAv,Dambis}).
     The resulting fluxes were corrected for the interstellar extinction. The value of
     the interstellar extinction $A_v = 5.4$ also was taken from \citet{KiminkiAv}.
     After this, the calculated spectra were convolved with the transmission curves of the
     standard U, B, V and R filters. Thus, we have calculated the grid of
     models with different luminosities and compared them with observations to
     derive the luminosity of the object.

       A recent study by \citet{binaryO11} has shown that \objeleven \ is a single-lined
       spectroscopic binary (SB1 type). Therefore, we decided to model \objeleven \
       as a single object. To determine the parameters of the \objeleven \ atmosphere we
       chose the model most resembling the observational spectrum from the calculated
       grid of models of O-stars \citep{me,me2013} and then started to refine the
       parameters of this model. We determined the effective temperature using the
       lines of He\,{\scriptsize II}~$\lambda4541.59, 5411.52$~\AA~ and  He\,{\scriptsize I}~$\lambda 4471.5,
       5875.66$~\AA,
       as well as the lines of nitrogen N{\scriptsize III}~$\lambda\lambda4634.0, 4640.6$~\AA~ \\ and
       weak absorptions of  N{\scriptsize IV}~$\lambda\lambda5200.60, 5204.28 $~\AA. In the spectrum
       \objeleven \ there are no lines of \\ NV~$\lambda\lambda4604.16,
       4620.5$~\AA \ that %%PA    ---------- which
       indicates that $T_{eff}$ is below 40000~K.

         The non-clumped mass-loss rate (${\dot{M}}_{uncl}$) is related to the clumped (${\dot{M}}_{cl}$) by the
         relation $\dot{M}_{uncl}=\dot{M}_{cl}/\sqrt{f_{\infty}}$, where $f_{\infty}$
         is the volume filling factor at infinity. We investigated the effect of clumping
         on intensities of lines.
         We have stopped at a value $f_{\infty}=0.08$ which well describes resonance
         lines of silicon Si{\scriptsize IV}~$\lambda\lambda1393.75, 1402.77 $~\AA \ and  $H_\alpha$ line.

\begin{figure*}[t]
%\begin{center}
%{\resizebox*{2.0\columnwidth}{!}{\includegraphics[angle=0]{uv_spect_obj11_aug11.eps}}}  % MODEL431 !!! {uv_spect_obj11_cont1.eps}
%{\resizebox*{2.0\columnwidth}{!}{\includegraphics[angle=0]{spect_obj11.eps}}}% MODEL441+448 !!! {uv_spect_obj11_cont1.eps}
%\vspace{3cm}
\includegraphics[angle=0,scale=0.4]{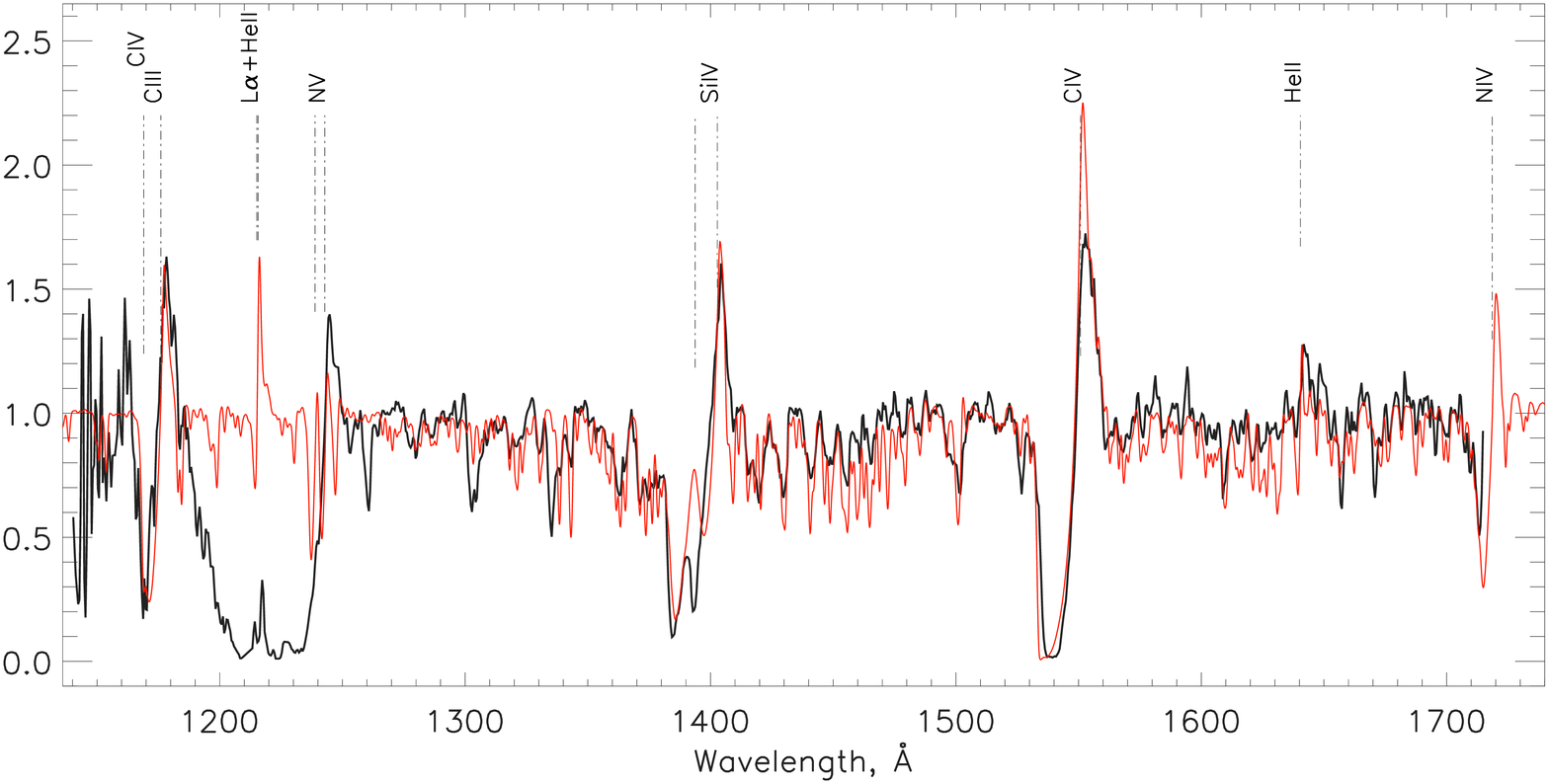}
\includegraphics[angle=0,scale=0.4]{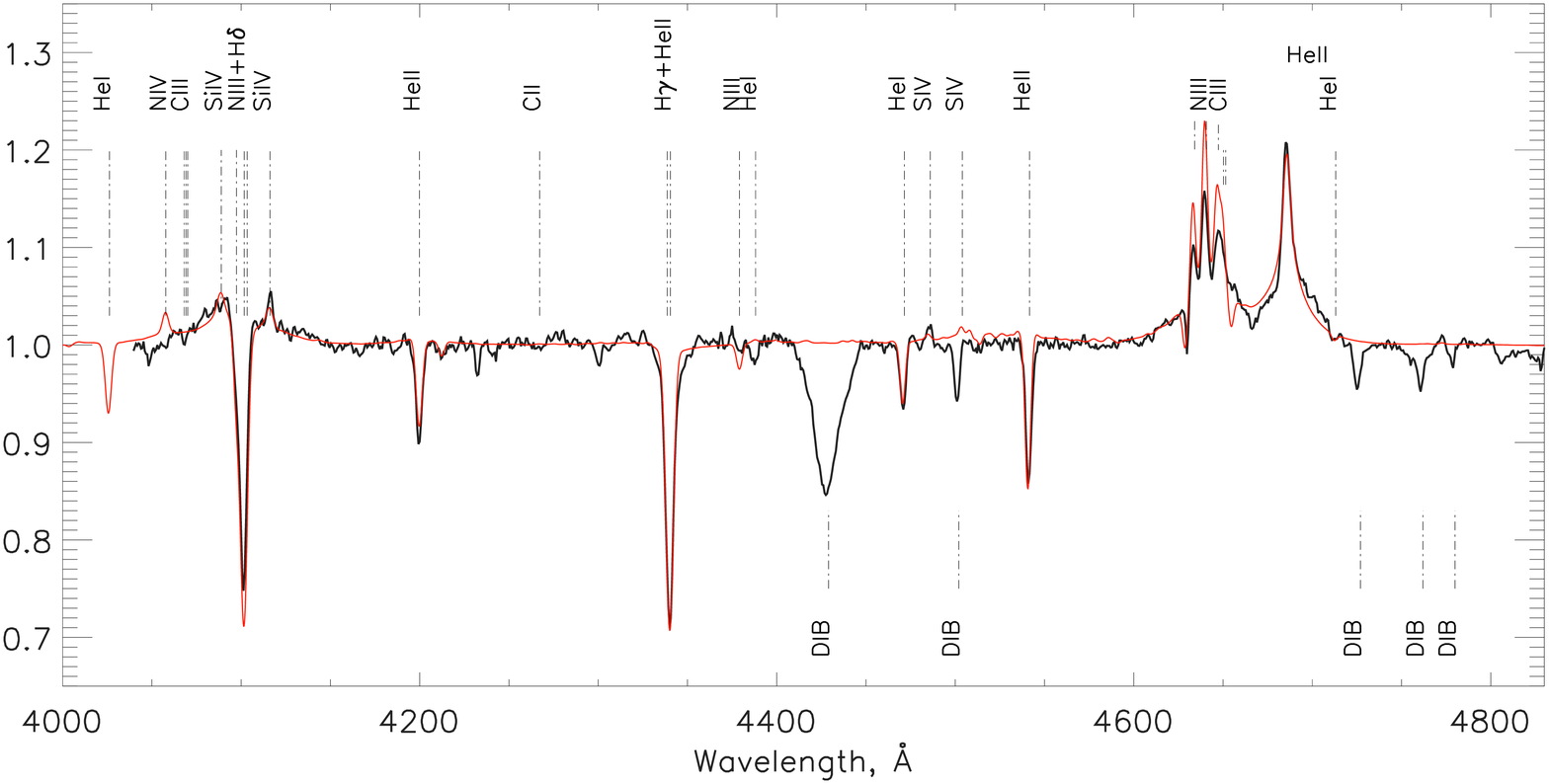}%{spect_obj11.eps}    MODEL458   !!!!!!!!!
\caption{Comparison of the observed spectrum (the black line) with the model (the red line).
         The top panel shows data obtained by the Hubble Space Telescope, while the bottom panel
         shows data obtained by the 4.2-m William Herschel Telescope.}
\label{fig:uvmodel11}
%\end{center}
\end{figure*}

\begin{figure*}[t]
  {\resizebox*{0.73\columnwidth}{!}{\includegraphics[angle=0]{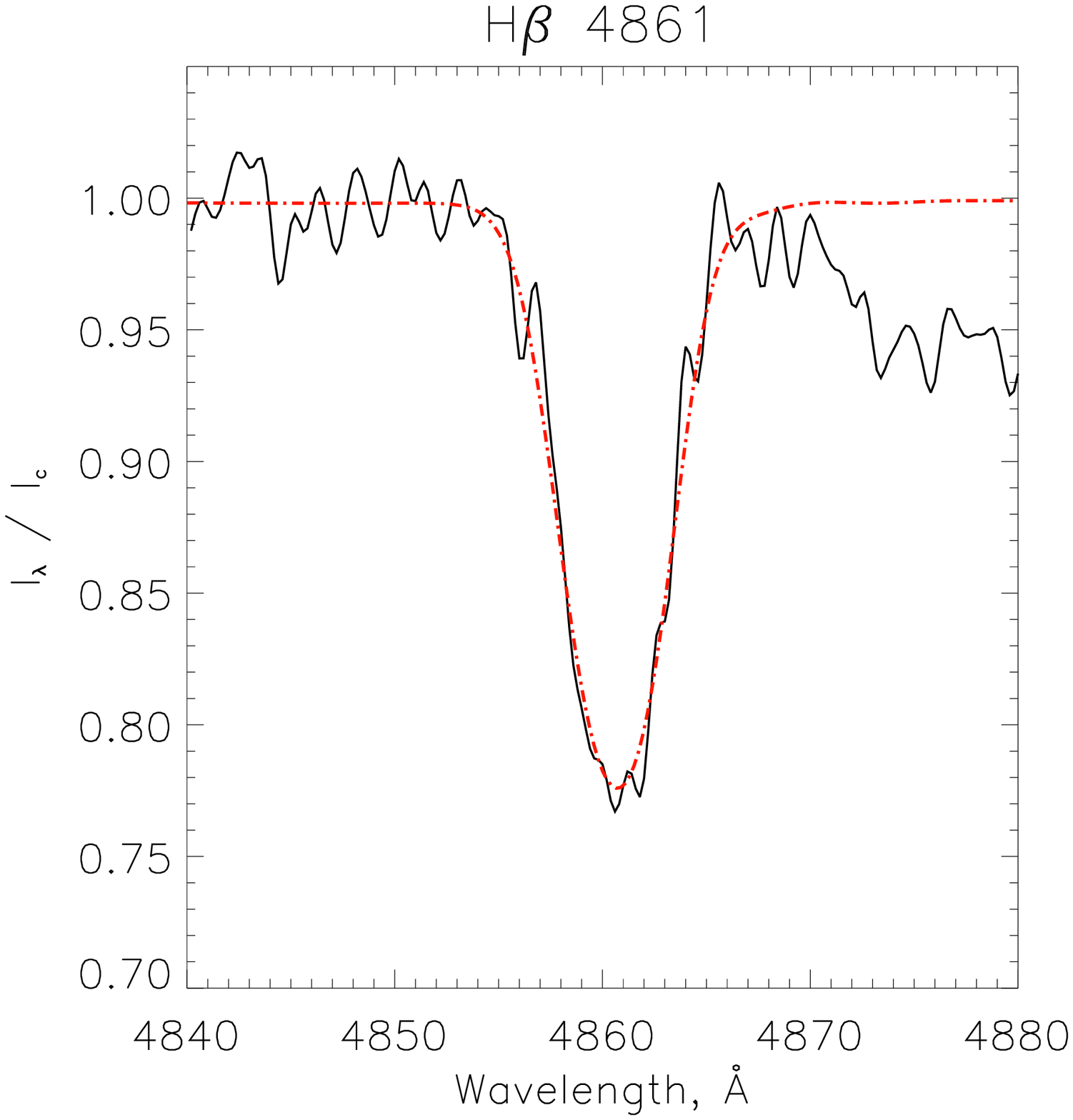}}}%{CygOB11_new_4861.eps}}}
  {\resizebox*{0.73\columnwidth}{!}{\includegraphics[angle=0]{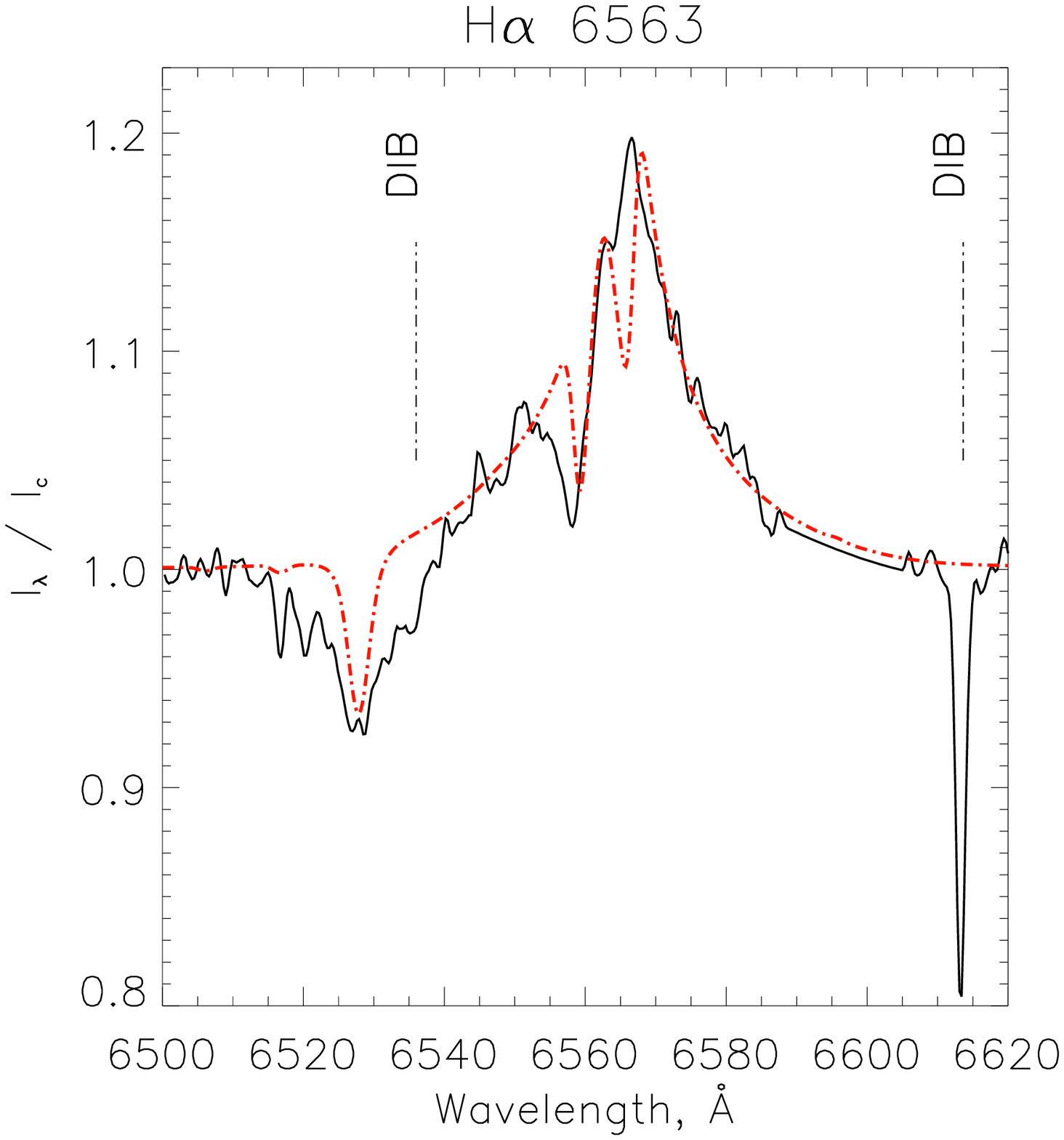}}}%{CygOB11_new_6562.eps}}}
  {\resizebox*{0.73\columnwidth}{!}{\includegraphics[angle=0]{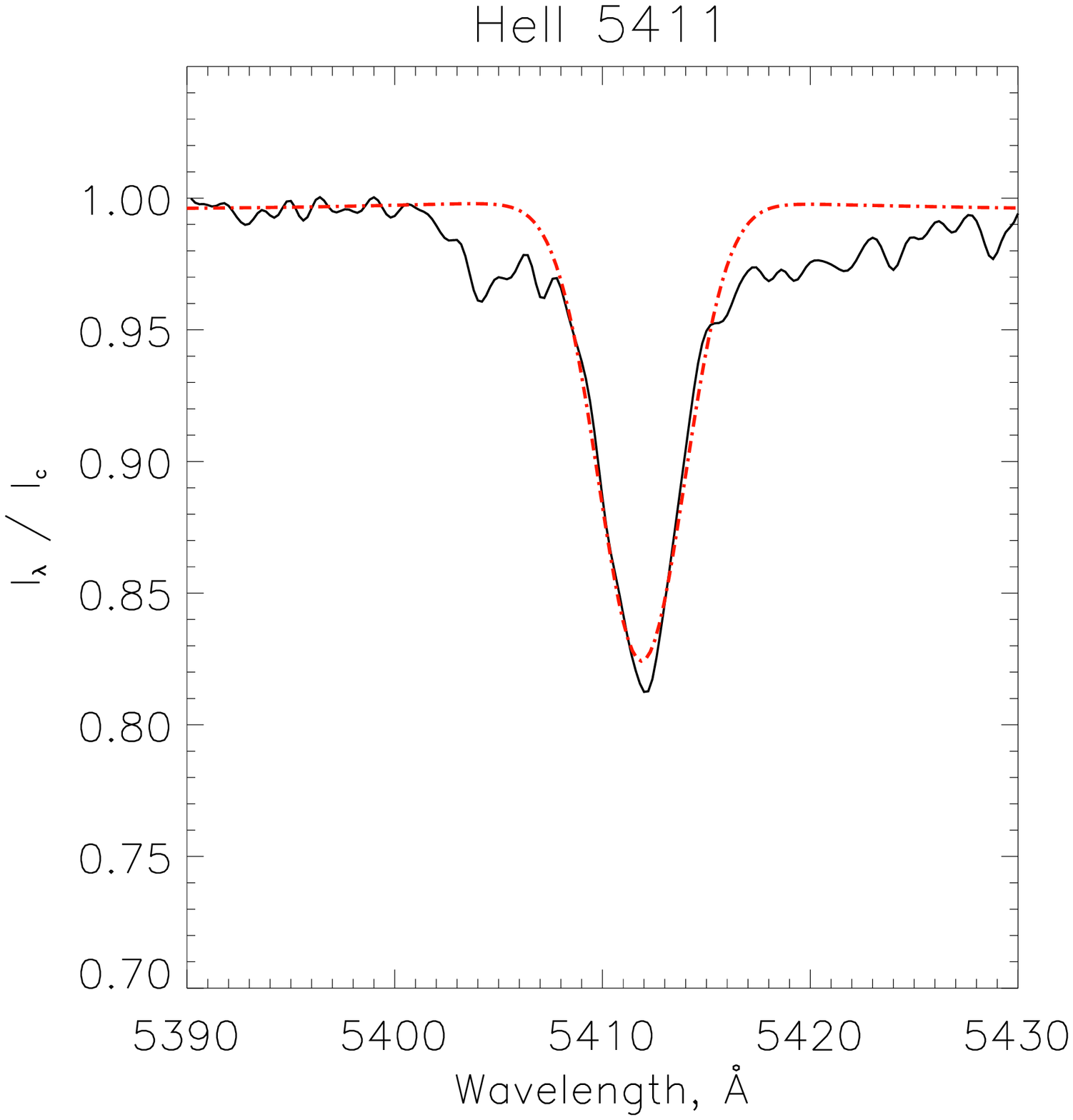}}}   %{CygOB11_new_5411.eps}}}
%%%%%%%%
  {\resizebox*{0.73\columnwidth}{!}{\includegraphics[angle=0]{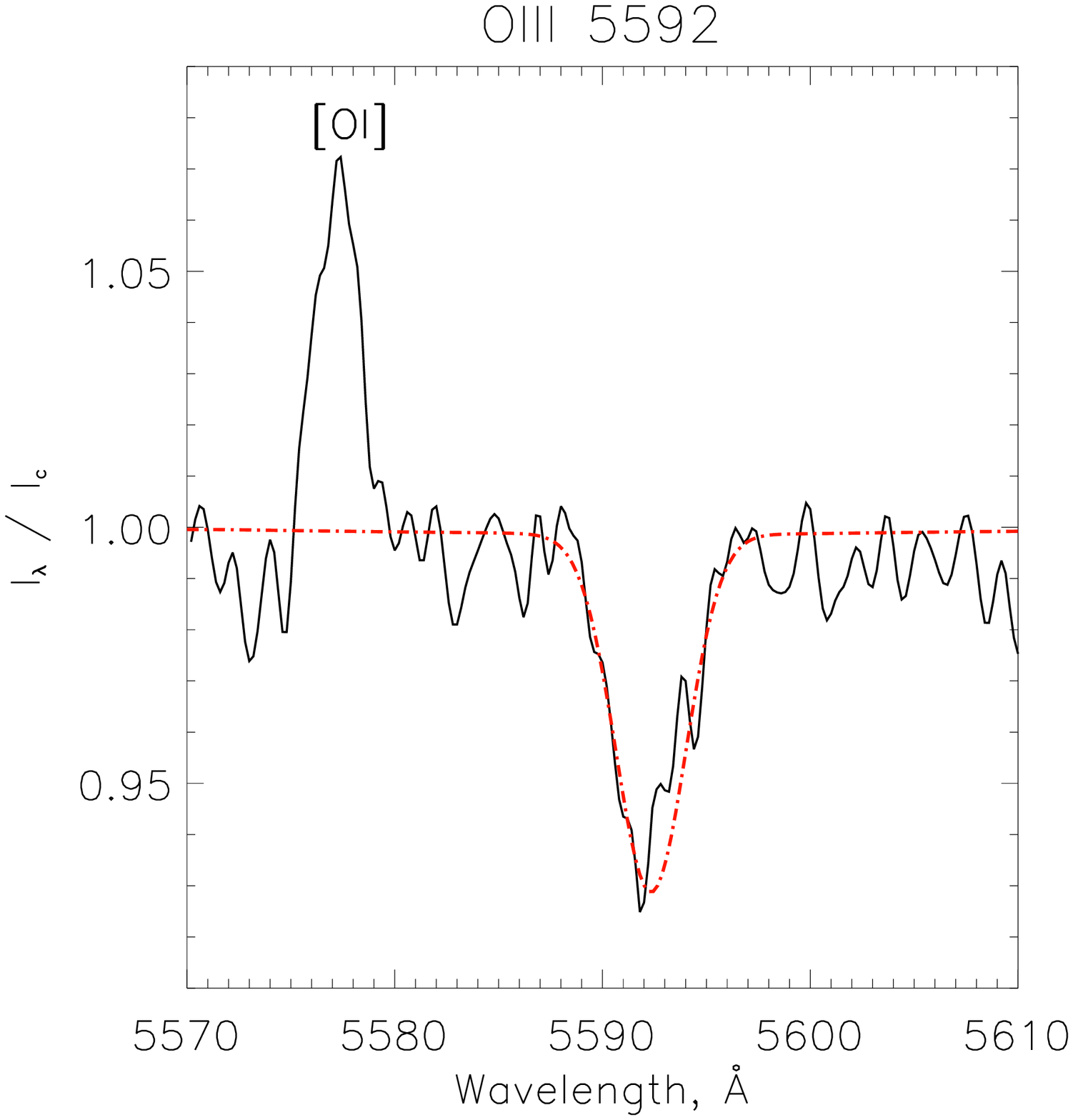}}}%{CygOB11_new_5592.eps}}}
  {\resizebox*{0.73\columnwidth}{!}{\includegraphics[angle=0]{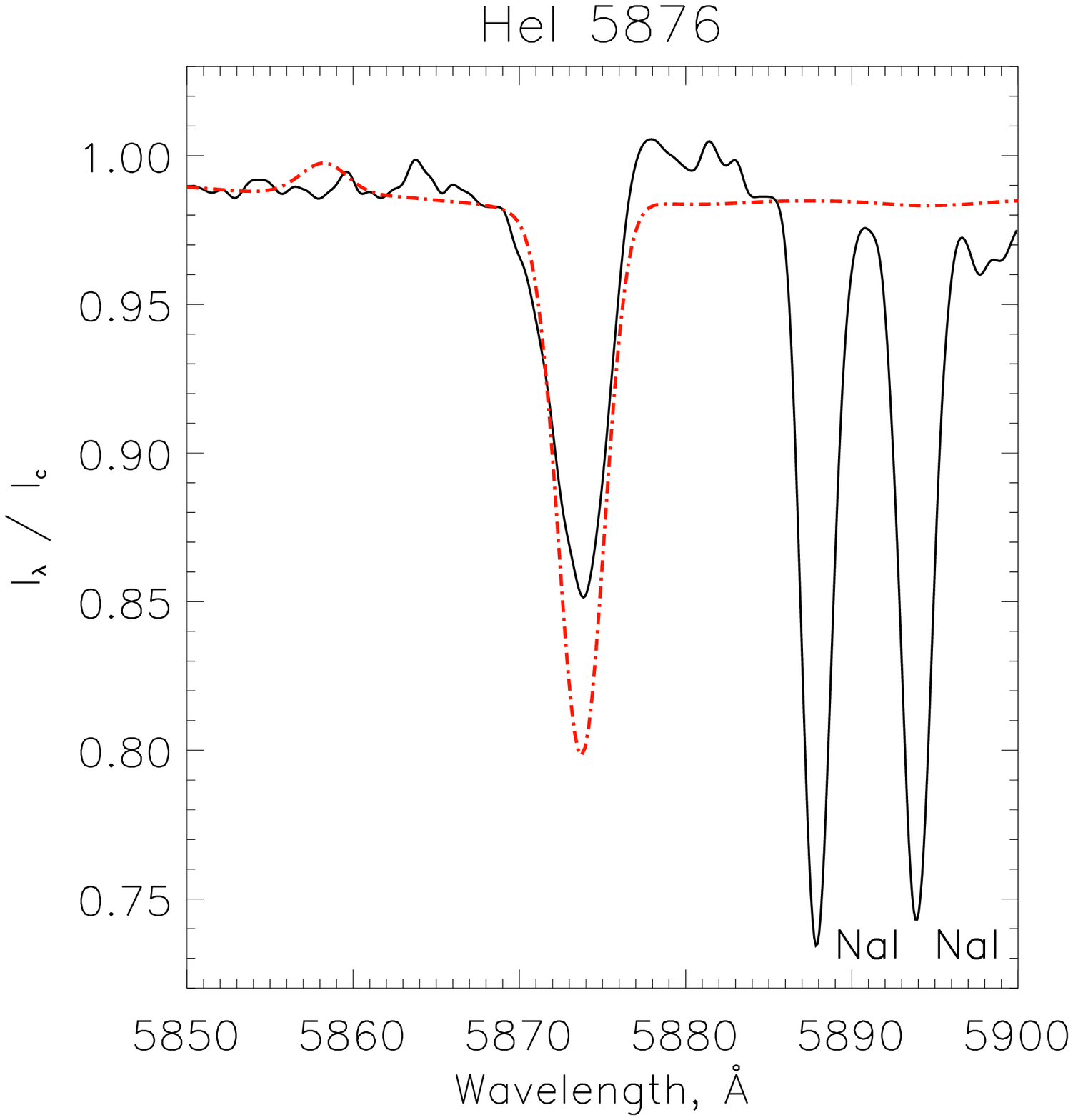}}}%{CygOB11_new_5876.eps}}}
  {\resizebox*{0.73\columnwidth}{!}{\includegraphics[angle=0]{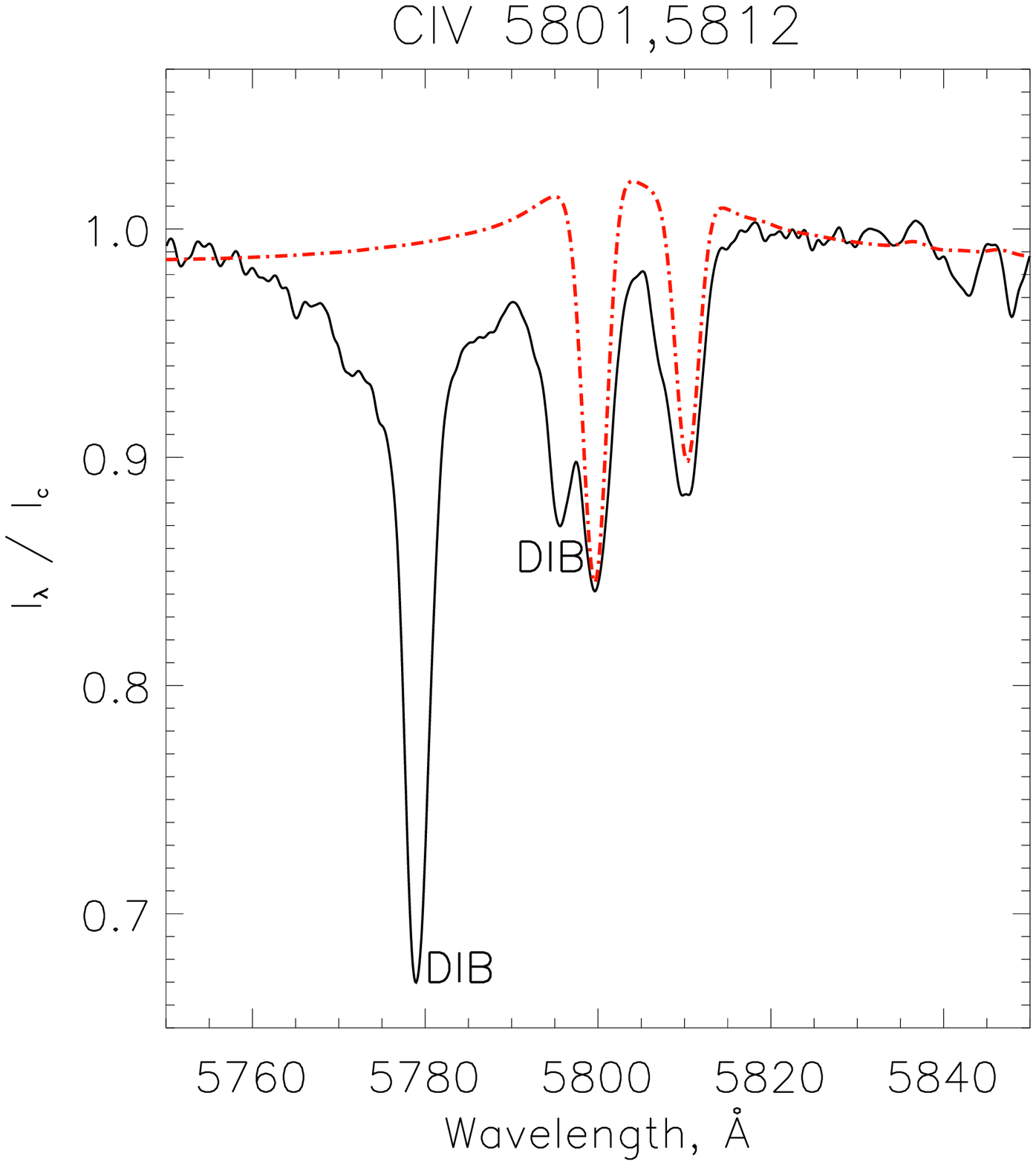}}}%{CygOB11_new_5880.eps}}}
%  {\resizebox*{0.66\columnwidth}{!}{\includegraphics[angle=0]{CygOB11_new_4340.eps}}}
%  {\resizebox*{0.66\columnwidth}{!}{\includegraphics[angle=0]{CygOB11_new_4541.eps}}}
%  {\resizebox*{0.66\columnwidth}{!}{\includegraphics[angle=0]{CygOB11_new_4686.eps}}}
%%%%%%
%  \vspace{3cm}
  \caption{Comparison of the profiles of selected lines with the best model spectra.
           The black line shows the observed profile, and the red line -- the model.}
  \label{fig:hbetamodel}
\end{figure*}

         Figures~\ref{fig:uvmodel11},~\ref{fig:hbetamodel}  present the comparison of  the
         observed spectrum of \objeleven \ with the best model in the optical and the ultraviolet region.
         For comparison, the Table~\ref{tab:measurements} gives the parameters of
         \objeleven \ derived by \citet{Herrero2002} and \citet{MokiemObj7}. %using the \sc fastwind} code \citep{fastwind,Puls}.
         The temperature difference is within error limits.
         Obvious difference in the luminosity is due to the difference in estimations of distance to the star.
         In the previous works distance modulus is suggested to be equal to $11.2\pm0.1$, which corresponds to  ${\rm \approx 1.7~\kpc }$.
         In the present work, as stated above, we assume that the distance is ${\rm 1.5~\kpc }$ \citep{Dambis}.

         Most probably the differences in the mass loss rate are due to taking clumping
         into account in our computations.
%%%%%%%%%%%%%%%%%%%%
        \citet{PulsMarkova} also have measured the mass loss rate taking clumping into account.
       They determined  that mass loss rate is  $\dot{M}_{cl}=(5\pm0.5)\cdot 10^{-6} \rm
       M_{\odot}\mbox{yr}^{-1}$ and $f=1$ for the region where $H_{\alpha}$ is formed. Therefore $\dot{M}_{cl}$ equal to $\dot{M}_{uncl}$ in this region.
       And our value of unclumped mass-loss rate is consistent with estimation \citet{PulsMarkova}.

       $\beta$ is the exponent in the velocity law describing the increase of velocity with radius.
       With the decrease of $\beta$  the depths of absorption lines vary and
       the  widths of %%PA
       He\,{\scriptsize II}~$4685.7$ and  $H_\alpha$ wings increase. In our calculations we stopped
       at $\beta=1.3\pm0.1$.
This value is higher than value of $\beta$-parameter obtained  for \objeleven
\  before (Table~\ref{tab:measurements}),
       but it is comparable to the $\beta$ parameters found for other O-class stars  using
       {\sc cmfgen} (Table~\ref{tab:startable}). %%PA

%%%%%%%%%%%%%%%%%%%%%%%%%%%%%%%%%%%%%%%%%%%%%%%%%%%%%%%%%%%%%%%%%%%%%%%%%%%%%%% MODEL431 !!!
\begin{table*}
\caption{Atmospheric parameters of \objeleven, derived in this work -- {\bf  (1)} and in the works of  \citet{Herrero2002} -- {\bf (2)} and   \citet{MokiemObj7} -- {\bf (3)}.}
\label{tab:measurements}
%\bigskip
\vspace{0.5cm}
\begin{tabular}{l|c|c|c|c|c|c|c|c}
%\hline
\label{tab:parmodels}
       &  $T_{eff}$,        &~$R_{2/3}$,~      &~$log L_*$,~  & ~$\dot{M}_{uncl}$,~                   & ~$\dot{M}_{cl}$,~                    &$f_\infty$ &  $V_{\infty}$,    &$\beta$\\
       &  kK                & $\rm R_{\odot}$  &   $L_{\odot}$  &$10^{-6}~\rm M_{\odot}\mbox{yr}^{-1}$& $10^{-6}~\rm M_{\odot}\mbox{yr}^{-1}$&           &      \kms         &        \\
\hline%&                    &                  &                &                                     &                                      &           &                   &        \\
%      &                    &                  &                &                                     &                                      &           &                   &        \\
       &                    &                  &                &                                     &                                      &           &                   &         \\
1    & $36^{+0.5}_{-1.0}$ &$20.7^{+2}_{-1.4}$&  $5.81\pm0.035$&     $6\pm0.7$                       & $1.7\pm0.2$                 & $0.8^{+0.2}_{-0.1}$&  2200             &$1.3\pm0.1$ \\
       &                    &                  &                &                                     &                                      &           &                   &         \\
2    & $37\pm1.5$         &  22.2            &  $5.89\pm0.05$ &     9.88                            &                                      &           &  2300             & 0.9     \\
3    &$36.5^{+0.4}_{-0.6}$&  22.1            &  $5.92\pm0.11$ &     7.36                            &                                      &           &  2300             & 1.03    \\

\end{tabular}
\end{table*}
%%%%%%%%%%%%%%%%%%%%%%%%%%%%%%%%%%%%%%%%%%%%%%%%%%%%%%%%%%%%%%%%%%%%%%%%%%%%%%% MODEL431 !!!

%888888888888888888888888888888888 figure 1 88888888888888888888888
%8888888888888888888888888888888888 figure 1 8888888888888888888888

%8888888888888888888888888888888888 figure 2 8888888888888888888888
%8888888888888888888888888888888888 figure 2 8888888888888888888888

\section{Chemical Composition}\label{sec:chemical}

       Calculated abundances of the basic elements are given in Table~\ref{tab:frac}.
       During the modeling we had difficulties with the description of He\,{\scriptsize II}~$4685.7$
       line.  This line is sensitive to changes in temperature %, and to changes of
       and in %%PA
       wind conditions -- clumping and mass-loss rate.
       At the outset of the modeling we  supposed that $\rm N_{He}/N_H=0.2$ (by number). % as for normal O-stars.
       We determined effective temperature through the He\,{\scriptsize II}
       to He\,{\scriptsize I} ratio and the mass loss rate through
       N{\scriptsize III}~$\lambda\lambda4634.0,  4640.6$~\AA \ and
       C{\scriptsize III}~$\lambda\lambda4647, 4650, 4652$~\AA \ lines, as well as $H_\alpha $ and
       $H_\beta$ lines in the way so $H_\alpha $ is in emission and $H_\beta$ in
       absorption, and as a result we got very strong He\,{\scriptsize II}~$4685.7 $ line. To decrease
       the intensity of this line we have significantly reduced the mass fraction of
       helium in the model down to $16-28$ \%, which is $0.05-0.1$ fraction by
       number of atoms. \citet{Herrero2002} determined the abundance of helium as 0.09
       (the He abundance by number of particles relative to H plus He). Our estimate
       agrees with this value within our error limits.
       The relatively minor He enhancement is consistent with characteristic of Ofc  spectral class \citep{WalbornHowarth}.

       Determination of carbon fraction in atmospheres of O-stars is a very important
       and difficult task. The surface chemical composition depends on the star's rotation
       rate, metallicity and mass. Surface chemical abundances %%PA
       are key to understand the  physical processes controlling the evolution of massive stars. In optical range there
       are C{\scriptsize IV}~$\lambda \lambda 5801.3, 5812$~\AA,
       C{\scriptsize III}~$\lambda\lambda4647, \\ 4650, 4652$~\AA~ and C{\scriptsize III}~$\lambda 5696$~\AA \ lines. But
       \citet{CIVMartins} do not recommend to use the lines of
       C{\scriptsize III}~$\lambda\lambda4647, 4650, 4652$~\AA~ and C{\scriptsize III}~$\lambda 5696$~\AA \ to
       determine the carbon abundance, because they are sensitive to $\log{g}$,
       effective temperature, mass loss rate, as well as to the inclusion of other
       ions in calculations, for example Fe{\scriptsize IV}, Fe{\scriptsize V}, S{\scriptsize IV} \citep{CIVMartins}.
       To determine the carbon abundance in the atmosphere the ultraviolet line C{\scriptsize III}~$\lambda1247$~\AA  \
       also may be used \citep{CIVMartins}.  But, as can be seen in \ Figure~\ref{fig:uvmodel11}, 
       C III 1247 is masked by strong N V emission so it can’t be used for abundance analysis.
%%%%%%%%
       We determined the carbon abundance in a spectrum of \objeleven \ using lines
       C{\scriptsize IV}~$\lambda\lambda 5801.3, 5812$~\AA \ and C{\scriptsize III}~$\lambda1175$~\AA
       \ (Fig.~\ref{fig:uvmodel11},~\ref{fig:hbetamodel}) 
       and we got ${\rm \epsilon(C)=12+log[N_C/N_H]=8.5\pm0.09}$. 
      Solar abundance of C is $8.39\pm0.05$ \citep{solarabundance}, so within the errors, the carbon abundance is solar.   %${\rm \epsilon(C)=8.4\pm0.2$,
       %${\rm 12+log[X/H]}=8.4\pm0.2}$       $X_C=0.35\pm0.05 \%$
       %i.e.  $ X_C/X_{C_\odot}\approx1.3$ by mass fraction.

       We estimated the nitrogen abundance primarily by emission lines of
       N{\scriptsize III}~$\lambda\lambda4634.0, 4640.6 $~\AA~ and weak absorption of
       N{\scriptsize IV}~$\lambda\lambda5200.60, 5204.28$~\AA. These lines are considered ideal for
       determining the nitrogen abundance \citep{OstarsBouret}. O{\scriptsize III}~$\lambda
       5592.25$~\AA~
       line is well distinguishable in the spectrum of the object. It has been used
       to estimate the oxygen abundance. %%PA

%8888888888888888888888888888888888 figure 3-0 8888888888888888888888
 \begin{figure*}[t]
  {\resizebox*{0.69\columnwidth}{!}{\includegraphics[angle=0]{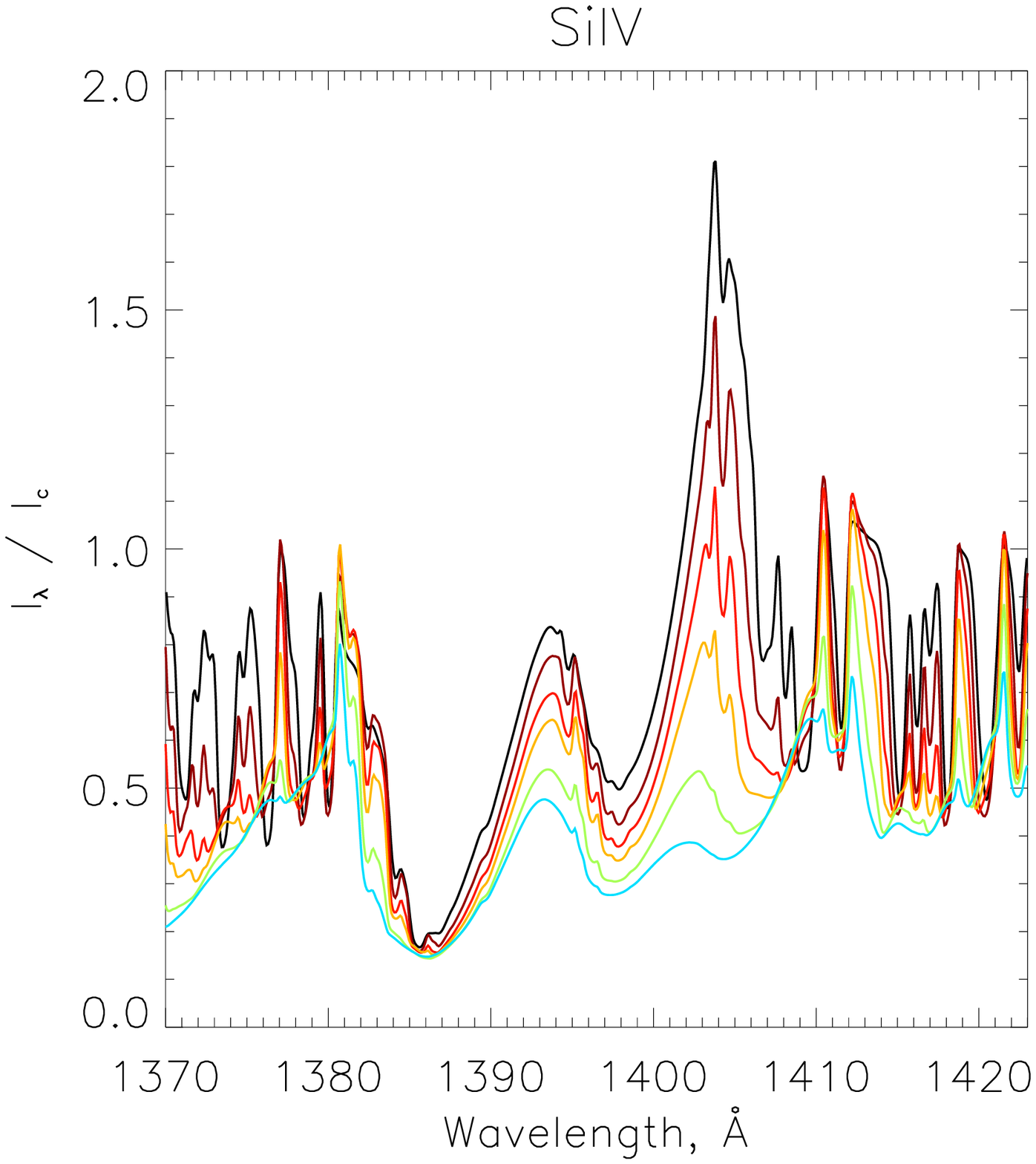}}}
  {\resizebox*{0.69\columnwidth}{!}{\includegraphics[angle=0]{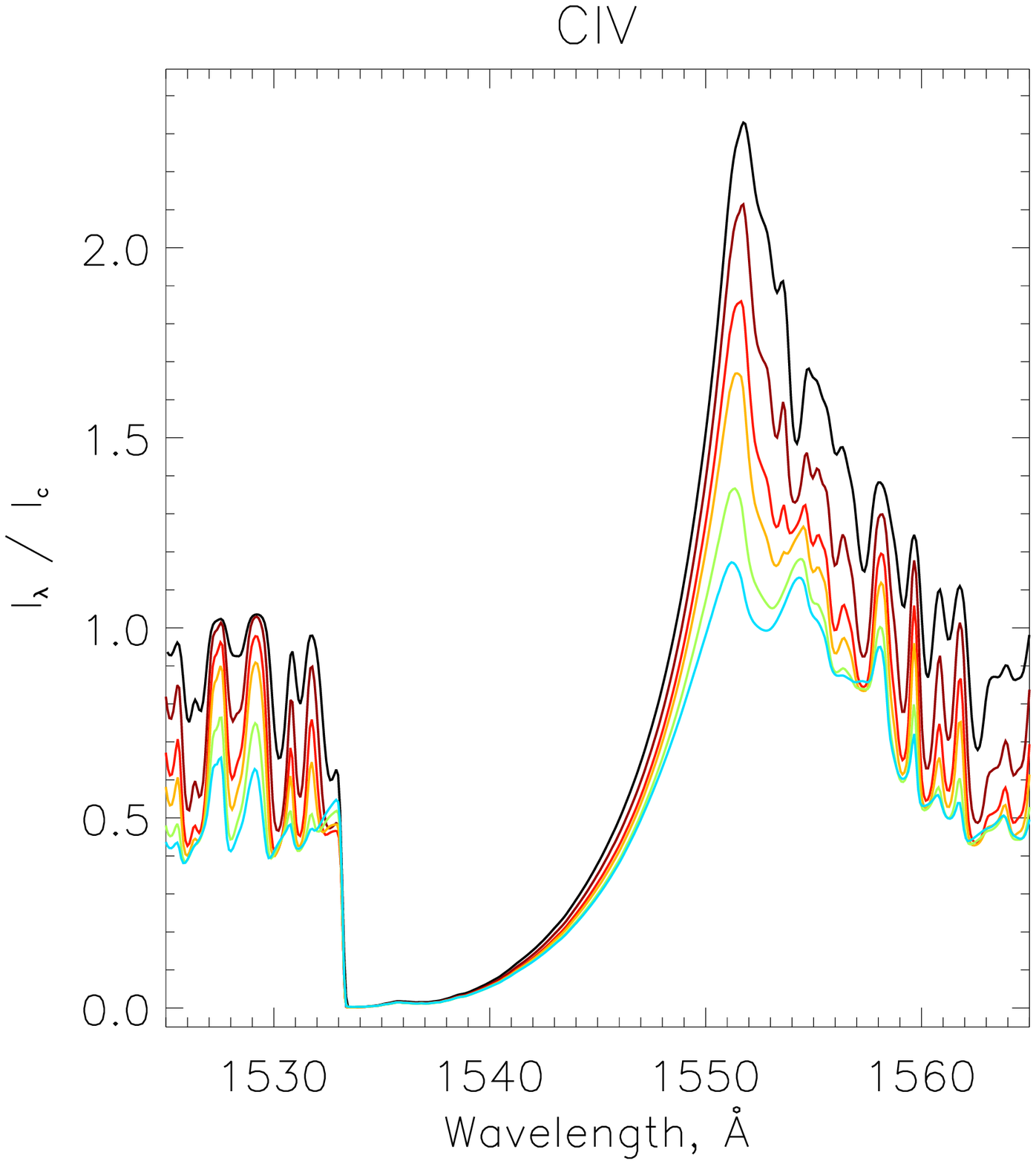}}}
  {\resizebox*{0.69\columnwidth}{!}{\includegraphics[angle=0]{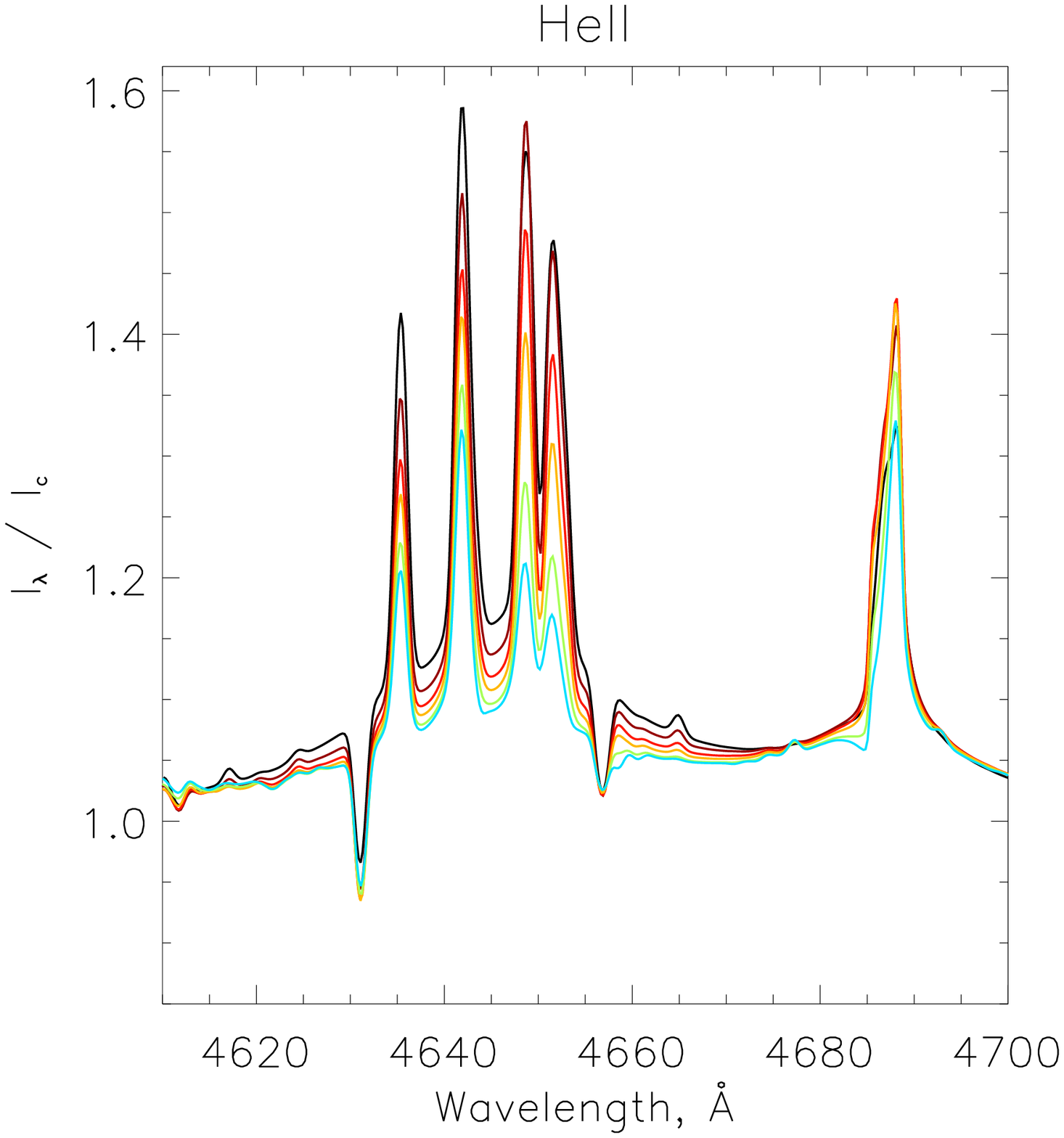}}}
%%%%%%%
  \caption{ The variation of the Si\,{\scriptsize IV}~$\lambda\lambda1393.75, 1402.77$~\AA,
            C\,{\scriptsize IV}~$\lambda\lambda1548.2, 1550.8$~\AA \ and He\,{\scriptsize II}~$4685.7$
            line profile depending on the iron abundance. Black line shows the model with
            ${\rm Fe_*/Fe_\odot }=0.37$, cherry -- ${\rm Fe_*/Fe_\odot }=1.5$, red -- ${\rm Fe_*/Fe_\odot }=3.7$,
            orange -- ${\rm Fe_*/Fe_\odot }=6.7$, green -- ${\rm Fe_*/Fe_\odot} =14.8$ and blue --
            ${\rm Fe_*/Fe_\odot }=24.3$. }
  \label{fig:Feline}
\end{figure*}
%8888888888888888888888888888888888 figure 3-0 8888888888888888888888

       In addition to lines of five basis elements H, He, C, N and O, lines of
       silicon and sulfur are present in the spectrum of \objeleven. We determined
       the abundance of silicon using weak absorption
       Si{\scriptsize IV}~$\lambda\lambda4629.85, \\ 4655.49$~\AA \ and emission
       Si{\scriptsize IV}~$\lambda\lambda6667.57, 6701.21$~\AA~ lines.  But we were not able to
       reproduce the doublet of Si{\scriptsize IV}~$\lambda\lambda1393.75, 1402.77$~\AA \ lines
       merging into the strong line with P~Cyg profile. This doublet increases with
       decreasing temperature, but we can not further decrease the temperature,
       because emission lines of He\,{\scriptsize I}  lines appear then  in optical
       range, %%PA appear --> should appear then ?
 for example,
       He\,{\scriptsize I}  ~$\lambda4921.94 $~\AA, which is not in the spectrum of the
       object. Silicon doublet Si{\scriptsize IV}~$\lambda\lambda1393.75, 1402.77$~\AA \ is also
       sensitive to the silicon abundance. But to get it, the silicon abundance in
       atmosphere of \objeleven \ should be seven times greater than solar one, which
       would significantly boost the Si{\scriptsize IV} optical lines that are also not observed
       in object spectrum.

 %%%%%%%%%%%%%% ÓÏÄÅÒÖÁÎÉÅ ÖÅÌÅÚÁ

We did not use  Fe {\scriptsize V} and Fe{\scriptsize VI} lines to determine the iron
abundance in the atmosphere of \objeleven \ because there are too many Fe lines in
ultraviolet region and these lines are blended into wide bands at low spectral resolution.
We used an indirect method instead. The resonance doublet of Si{\scriptsize IV}~$\lambda\lambda1393.75,
1402.77$~\AA \ is sensitive to many factors, including Fe abundance. It becomes weaker when
iron content  in the atmosphere  increases (see Fig.~\ref{fig:Feline}). We found that
the Si{\scriptsize IV}~$\lambda\lambda1393.75, \\ 1402.77 $~\AA \ lines in the spectrum of
\objeleven \ may be described if we significantly reduce (down to 0.37~${\rm X_{Fe{_\odot}}}$)
the abundance of iron in the object atmosphere.  %%PA: what are the possible
%%reasons or implications of this low metallicity?

%%%%%%%%%%%%%%%%%%%%%%%%%%%%%%%%%%%%%%%%%%%%%%%%%%%%%%%% MODEL 403
\begin{table*}[h]
\begin{center}
\caption{The abundances of chemical elements are given in the logarithmic scale relative hydrogen -- $\rm log(N_{el}/N_H)+12$, where $N_{el}$ is the abundance of a given element by number.}
 %X$_i$/X$_\odot$ --  is the ratio of the element abundance to the solar value. }
\vspace{0.5cm}
\label{tab:frac}
\begin{tabular}{lll}%||lcc}
%          &                        \\ 
 Element   &   \objeleven       &   Sun           \\        %& $\epsilon(X)$          \\       
\hline %   &                    &                 \\   
     H     &    12              &  12             \\%$1.0$               \\
     He    &    $10.85\pm0.15$  &  $10.93\pm0.01$ \\%$0.075\pm0.025$                    \\
     C     &    $8.5  \pm0.09$  &  $8.39\pm0.05$  \\%$(3.3\pm0.7)\cdot 10^{-4}$          \\
     N     &    $8.28 \pm0.03$  &  $7.78\pm0.06$  \\%$(1.9\pm0.15)\cdot 10^{-4}$         \\
     O     &    $8.17 \pm0.07$  &  $8.66\pm0.05$  \\%$(1.5\pm0.25)\cdot 10^{-4}$         \\
     Fe    &    $7.14 \pm0.09$  &  $7.45\pm0.05$  \\%$(1.4\pm0.3)\cdot 10^{-5}$         \\
    Si     &    $7.81 \pm0.07$  &  $7.51\pm0.04$  \\%$(6.5\pm1)\cdot 10^{-5}$        \\
     S     &    $6.54 \pm0.06$  &  $7.14\pm0.05$  \\%$(3.5\pm0.5)\cdot 10^{-6}$        \\
\hline
\end{tabular}
\flushleft{The solar abundances are taken from \citet{solarabundance}} \\
\end{center}
\end{table*}
%%%%%%%%%%%%%%%%%%%%%%%%%%%%%%%%%%%%%%%%%%%%%%%%%%%%%%%%

 %%%%%%%%%%%%%%%%%%%%%%%%%%%%%%%%%%%%%%%%%%%%%%%%%%%%%%%%%%%%%%%%%%%%%%%%%%%%%%
\begin{table*}[h]
\begin{center}
\caption{Comparison of the atmospheric parameters of \objeleven \ with parameters of other O-stars.}
\label{tab:startable}
\begin{tabular}{llllcccccc}%{l|c|c|c|c|c|c|c|c}
%\hline
                           &             &          &                    &                             &           &                &         &         &     \\
                           &Sp. type$^*$ &$T_{eff}$,&~~$L_*$,~~          & ~~$\dot{M}_{cl}$,~~         &$f_\infty$ &  $V_{\infty}$, &$V~sin I$& $\beta$ &  Ref. \\
                           &             &  kK      &$10^5~\rm L_{\odot}$&$\rm M_{\odot}\mbox{yr}^{-1}$&           &     \kms       &   \kms  &         &    \\
\hline  %                  &             &          &                   &                              &           &                &         &         &    \\
%                          &             &          &                   &                              &           &                &         &         & \\
\footnotesize{Cyg OB2} \#11& O5 Ifc      & 36       &  6.5              & 1.7e-6                       &   0.08    &  2200          & 120     &   1.3   &    \\
                           &             &          &                   &                              &           &                &         &         &    \\
%Cyg OB2 \#8C&             & 41.8        &  4.9     &     3.37                           &             &  2650         &   & 0.85                              &    \\
%Cyg OB2 \#8A &            & 38.2        &  12.6    &     10.4                           &             &  2650         &   & 0.74                             &     \\
%             &            &             &          &                                    &             &               &   &                            &     \\
\footnotesize{Cyg OB2} \#8C& O4.5 III(fc)& 37.4     &  3.63             &     2.0e-6                   &   0.1      &  2800         & 175     & 1.3     & a    \\
\footnotesize{Cyg OB2} \#8A& O5 III(fc)  & 37.6     &  13.2             &     3.4e-6                   &   0.01     &  2700         & 110     & 1.1     & a    \\
210839                     & O6.5 If     & 36       &  6.3              &     1.41e-6                  &   0.05     &  2100         & 210     &   1.    & b    \\
163758                     & O6 I(n)fp   & 34.5     &  5.75             &     1.59e-6                  &   0.05     &  2100         & 94      &   1.1   & b  \\
15570                      & O4 If       & 38       &  8.7              &     2.19e-6                  &   0.05     &  2200         & 97      &   1.1   & b   \\
14947                      & O4.5 If     & 37       &  6.76             &     1.41e-6                  &   0.03     &  2300         & 130     &   1.3   & b   \\
Obj14                      &  O5/6 If    & 32.5     &  5.8              &     0.37e-6                  &            &  300          &         &   1.0   & c   \\
Obj3                       &  O5/6 If    & 30       &  3.7              &     2.71e-6                  &            &  800          &         &   0.8   &  c  \\
\hline
\end{tabular}
\flushleft
{$^*$ --  Spectral classes are listed according to \citet{Sota} catalogue.} \\
{  a --\citet{NajarroLband},  b --\citet{OstarsBouret},    c -- \citet{Borisova}.} \\
{  $V~sin I$ for \objeleven \ are taken from  \citet{HerreroUV}.}\\
\end{center}
\end{table*}
%%%%%%%%%%%%%%%%%%%%%%%%%%%%%%%%%%%%%%%%%%%%%%%%%%%%%%%%%%%%%%%%%%%%%%%%%%%%%%

%888888888888888888888888888888888 figure 3 88888888888888888888888
\begin{figure*}[h]
\begin{center}
\includegraphics[scale=0.7]{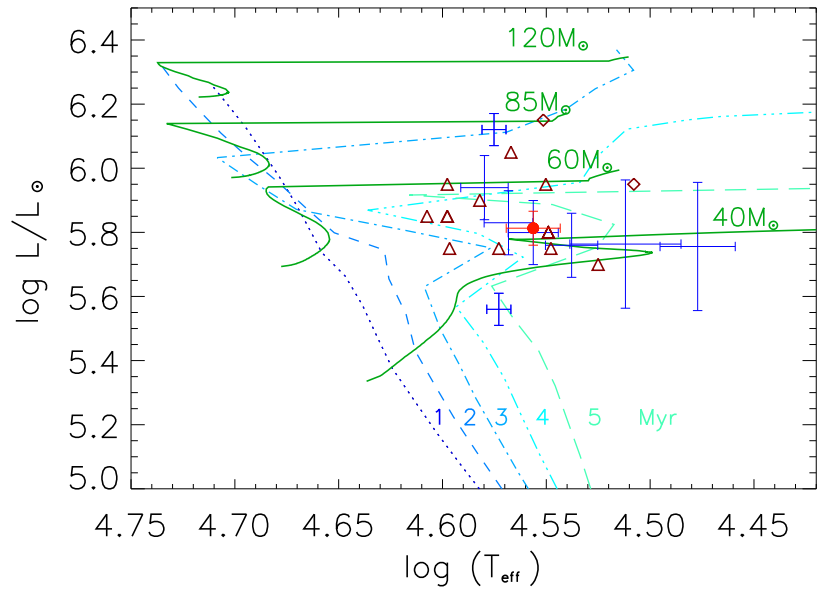}
%\vspace{3cm}
\caption{The location of \objeleven \ in the Hertzsprung-Russell diagram (marked with the red circle with  error bars).
         For comparison the stars of similar spectral class  are shown. Blue points with error bars show stars
         listed in Table~\ref{tab:startable}. Triangles are ``normal'' O4-6 supergiants belonging to  Arches cluster,
         diamonds are ``extreme'' O4-6~If$^+$ belonging to Arches cluster. These data was  taken from  \citet{MartinsOsupergiant}.
         The solid horizontal lines represent the mass tracks for stellar masses 120, 85, 60 and 40~\Msun. The vertical solid
         lines represent stellar isochrones. The evolution tracks and the stellar isochrones are taken from the Geneva library.}
\label{fig:HRdiagram}
\end{center}
\end{figure*}
%8888888888888888888888888888888888 figure 3 8888888888888888888888

%888888888888888888888888888888888 figure 4 88888888888888888888888
\begin{figure*}[h]
\begin{center}
\includegraphics[scale=0.7]{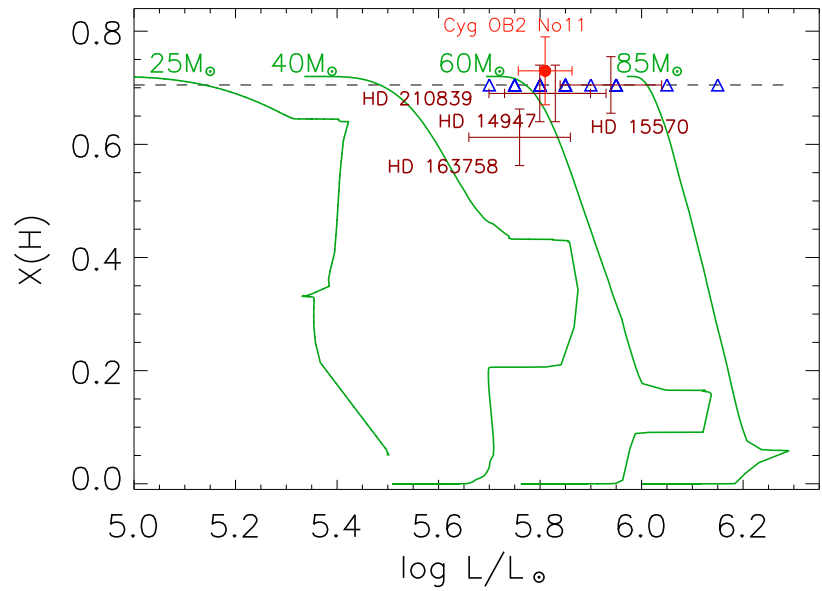}
%\vspace{3cm}
\caption{Hydrogen mass-fraction as a function of luminosity. \objeleven \ are
  marked with the red circle with error bars.  Locations of O4-6 and
  O4-6~If$^+$ supergiants belonging to Arches cluster are marked by
  triangles. Solid lines show evolution tracks from Geneva library. }
\label{fig:diagramHyd}
\end{center}
\end{figure*}
%8888888888888888888888888888888888 figure 4 8888888888888888888888
%888888888888888888888888888888888 figure 5 88888888888888888888888
\begin{figure*}[h]
\begin{center}
\includegraphics[scale=0.7]{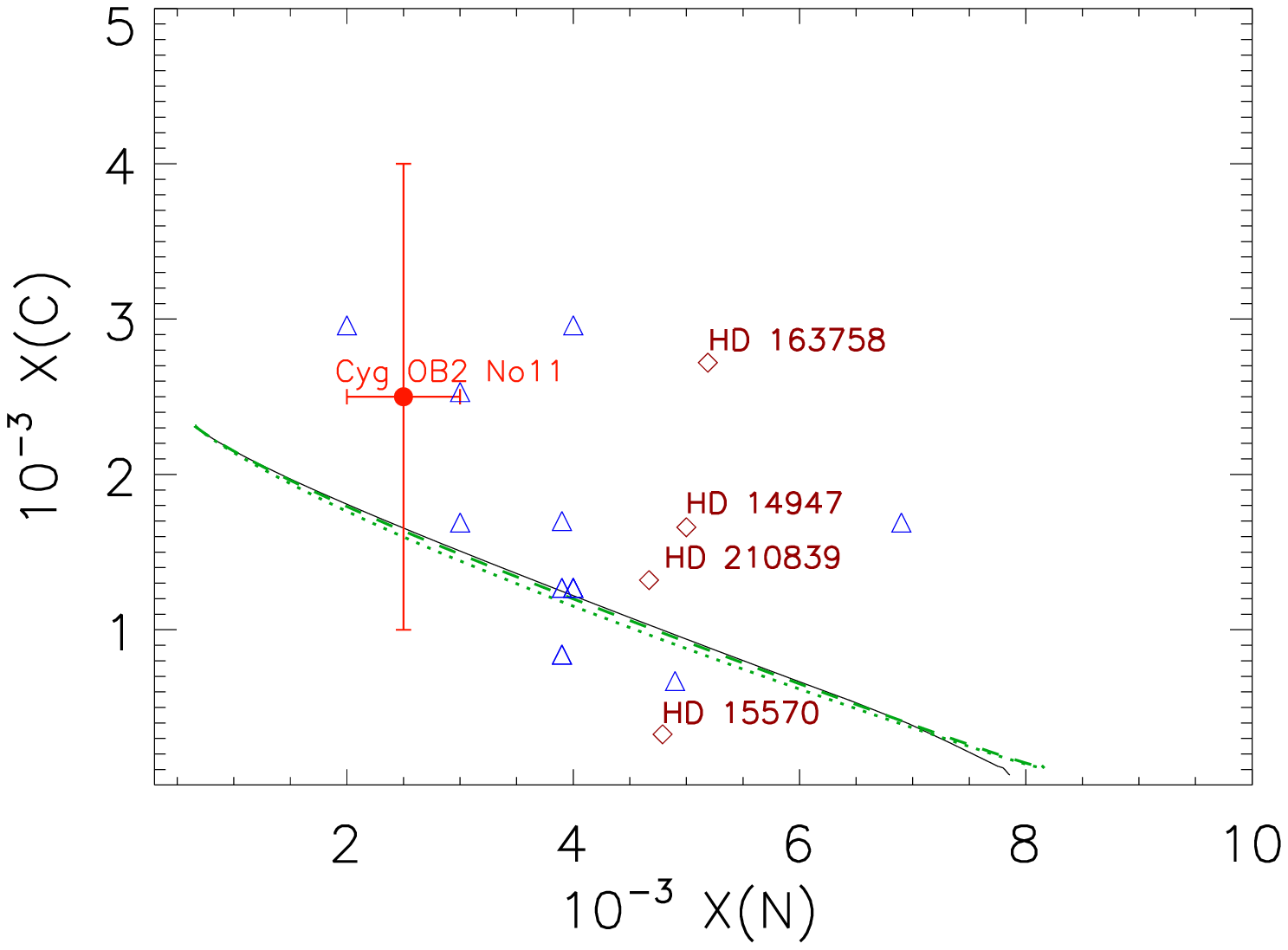}
%\vspace{3cm}
\caption{Carbon mass-fraction as function of nitrogen mass-fraction, as
  predicted by Geneva model with the rotation rate $V_{rot}/V_{crit}=0.4$.  %an initial rotation velocity .
Solid
  line is for 40~\Msun, dashed for 60~\Msun, dotted for 85~\Msun. Unnamed
  triangles mark position of O4-6 and O4-6~If$^+$ supergiants belonging to Arches
  cluster.}
\label{fig:diagramNit}
\end{center}
\end{figure*}
%8888888888888888888888888888888888 figure 5 8888888888888888888888

\section{Locations  of {\objeleven} on the Hertzsprung-Russell  Diagram}\label{sec:diagram}
%%%%%%%%%%%%%%%%%%%%%%%%%%%%%%%%%%%%%%%%%%%%%%%%%%

       In Table~\ref{tab:startable} the parameters of \objeleven \ are compared with the
       parameters of other O-stars of similar spectral
       classes. Figure~\ref{fig:HRdiagram} shows the locations of these stars on the
       Hertzsprung-Russell diagram as well as evolutionary tracks and isochrones from
       the Geneva database \citep{Ekstrom}, constructed using the online calculator\footnote{
       http://obswww.unige.ch/Recherche/evol/-Database-}.
       The evolutionary tracks and isochrones are computed taking into account the effects of rotation. The rotation rate is $v_{ini}/v_{crit}=0.4$.
       It can be seen that
       \objeleven \ is closest to star HD163758 (O6 I(n)fp) and HD210839 (O6.5 If) and
       it is sufficiently far from two other O(fc) stars {\rm Cyg~OB2~\#8~C} %{\rm Cyg~OB2~\textnumero8~C}
       and {\rm Cyg~OB2~\#8~A} that are also marked on the diagram. The %%PA
       figure shows that the object lies between the isochrones corresponding to four
       and five Myr and the mass of the object is 40-60~\Msun.
       \citet{Chentsov} spectroscopically confirmed age differences for stars which are located in
       different parts of Cyg~OB2. Stars located at north part  of the  association are older than
       all other objcts in the Cyg~OB2. Their ages are about 5~Myr \citep{Chentsov}.
       The results of our numerical simulations support this conclusion and
       are consistent with the hypothesis of  cascade  star formation in the association Cyg~OB2      \citep{mearxiv}.
% ссылка на архив!!!!!!!!!!

       Figure~\ref{fig:diagramHyd} shows the change of hydrogen mass fraction with
       age. According to this diagram the mass of \objeleven \ is 60~\Msun. The figure shows
       that the object, like HD15570, is on the early part of its evolution track unlike
       stars HD163758 (O6.5~If) and HD210839 (O6~I(n)fp). It should be noted that
       although strong C{\scriptsize III} \ lines %~$\lambda\lambda4647-4650-4652$~\AA
       \ are observed in the
       spectra of HD15570 and HD14947, they are weaker  than the N{\scriptsize III} \
       lines, and these stars are not Ofc stars \citep{WalbornOIfc}. %%PA:
       %%why? what criterion fails?

       Consider the location of \objeleven \ on the diagram of carbon mass-fraction as
       function of nitrogen mass-fraction (Figure~\ref{fig:diagramNit}). This figure
       shows the evolutionary tracks for stars with masses of 40~\Msun, 60~\Msun and
       85~\Msun, computed in 2013 \citep{Ekstrom}. \citet{Maeder2009} write that N/C
       ratio is sensitive to many parameters such as the age of the star, metallicity,
       rotation velocity, magnetic field. Evolutionary tracks computed in 2013  are
       below the tracks designed in 2003 for the same stellar masses
       \citep{Maeder2003}. \objeleven \ is located above the evolutionary tracks.
       While considering the carbon abundance there are some objects on this diagram
       similar to \objeleven,
       %this object significantly differs from all the other O4-6 supergiants in nitrogen abundance.
        but as Figure~\ref{fig:diagramNit} shows, the nitrogen  abundance in this 
       object  is lower than that of other O4-6 supergiants.
       The abundance $\epsilon(N)=8.28\pm0.03$ obtained
       by us is lower than one for ``normal'' O4-6 supergiants.  Also it is lower than the
       nitrogen abundance in other supergiant belonging to Cyg~OB2, \objsev \
       (${\rm O3~If_*}$, $\epsilon(N)=8.65\pm0.05$) \citep{me2013}. 
       \citet{WalbornLMC} suggested the hypothesis that the morphologically normal majority of OB supergiants may be
       nitrogen enhanced, while the OBC minority have normal CNO abundances and the OBN class displays more
       extreme degrees of processed material mixed into the atmospheres. In general, our result is consistent with the  hypothesis.   
       Based on the calculations we conclude that nitrogen abundance in OBC stars can be up to two times greater than solar.

%%%%%%%%%%%%%%%%%%%%%%%%%%%%%%%%%%%%%%%%%%%%%%%%%%
\section{Search for the binary companion of  \objeleven}\label{sec:binary}

     As mentioned above, \objeleven \ is a SB1 binary star
     \citep{binaryO11}, with period of $72.43\pm0.07 $ days and $e=0.5\pm0.06$.
     It is known that the fraction of binary and multiple stars among young
     early-type objects is about 80\% or more that is more than twice above
     that for old low-mass stars. %%PA
     Moreover, most probable multiplicity of
     a newly forming stellar system is  $2-3$ \citep{good05}.
     Thus, even among the field stars the fraction of the triples and systems of higher multiplicity
     is approximately 0.25 of the total number of binaries.  All this makes it
     reasonable to search for the companions of objects like \objeleven.
     The search for companions of \objeleven \ was already performed by \citep{Apellaniz}.
     \citet{Apellaniz} has not found companions in the range of separations of $0.1-14 \ \arcsec $
     and the magnitude differences lower than $8~\mag $.

     So, we performed observations on 6-meter Russian BTA telescope with the speckle  interferometer, mounted in
     the primary focus.
%This device led to detect multiplicity up to the BTA difraction limit, which
%     is  $0.02$ \ \arcsec.
    This device allows to raise the angular resolution up to the diffraction
    limit which is  $0.02$ \ \arcsec\ for BTA. %%PA

     Observations with  speckle interferometer  based  on EMCCD
     \citep{maks2009} were carried out in the visual spectral range
     with filters having central wavelengths of 5500 and 8000~\AA \
     and pass-band halfwidths of 200 and 1000~\AA, respectively.
%     (effectively wavelength is 5450~\AA and 8000~\AA, respectively) under moderate
%     wheather conditions with seeing about $2 \  \arcsec$.
We estimated the atmospheric seeing from the full width at half maximum averaged over a
sequence of speckle frames for a stellar image, and found it to be about $2 \  \arcsec$.

     2000 frames of speckle images were accumulated in each series, with single
     frame exposure of  20 \,ms.
     Methods of reduction of speckle interferometric data from BTA is described in the articles \citet{maks2009,bal2002}.
     Method of the companion detection is based on analysis of the power spectrum averaged over sequence of speckle interferograms.
     We report the non-detection of a components within separation range of $0.02-4 \ \arcsec$ and with magnitude
     difference less than $6~\mag$ at given wavelengths. Indeed, \objeleven \ is extremely bright, so, the chance
     to find a companion that is 250 times weaker is little (but not zero!).  Let us mention the fact that at
     \objeleven's distance (1.51 \kpc \ according to \citep{KiminkiAv}) $0.02$ \ \arcsec \ separation corresponds to 30 AU.
     So, there is no {\bf wide} component in this system. And to fill the gap between the direct components detection
     at the largest telescopes diffraction limit and ability to SB detection we need to improve angular resolution
     down to at least  0.001 \ \arcsec. The search may be continued in IR range when the instrumental opportunities
     of high $\Delta \mag$ detection will be improved.

%%%%%%%%%%%%%%%%%%%%%%%%%%%%%%%%%%%%%%%%%%%%%%%%%%
\section{Results}\label{sec:results}

     In this work we investigated the atmosphere of ${\rm O5~Ifc} $ supergiant \objeleven \ using the
     medium-resolution spectra, obtained at RTT150 and WHT, and archival UV data.  We have applied non-LTE
     {\sc cmfgen} code for modeling of the atmosphere and determined its physical parameters and chemical composition.
     Parameters of \objeleven \ are similar to the ones of other O4-6 supergiants.  The position
     on the Hertzsprung-Russell diagram corresponds to the mass of star about 50~\Msun \ and age about
     4.5 Myr. Our estimation of age is consistent with ages of other stars located at north part of Cyg~OB2.

     We found that in the atmosphere of stars $He/H\approx0.1$ and nitrogen abundance is lower
     than that for other ``normal'' O stars ($\epsilon(N)=8.28\pm0.03$), while the carbon abundance is solar. 
     Overall, our modeling confirms
     the hypothesis sugested by \citet{Walborn76,WalbornLMC}, that   OBC supergints have normal CNO abundances
     and anomalies in N versus  C, O  correlate with the He abundance \citep{WalbornHowarth}. 
%     Based on our calculations we conclude that nitrogen abundance in OBC stars can be up to two times greater than solar, that increases the possible interval of N abundances in these stars.
     In the spectrum of  \objeleven \ there are lines of silicon and sulphur, which were used for estimating the
     abundances of these elements, as results   $\epsilon(Si)=7.81\pm0.07$ and $\epsilon(S)=6.54\pm0.06$.

     The speckle interferometry performed on the Russian 6-m BTA telescope does
     not reveal any binary companion in the
     range of separations of $0.02-4 \ \arcsec $  and with magnitude differences less than $6~\mag $.

%\acknowledgements
\section*{Acknowledgments}
       We would like to thank John D. Hillier for his great code {\sc cmfgen}.
       Likewise, we thank Nolan Walborn for valuable discussions and for providing
       the data obtained at WHT.  We would like to thank  the anonymous referee for
       valuable comments. We used the data from the Multimission Archive at STScl (MAST),
       the CASU Astronomical Data Centre, SIMBAD database and database of stellar evolution
       group at the Geneva Observatory. The study was supported by the Russian Foundation
       for Basic Research (projects no. {14-02-31247}, 12-07-00739, 12-02-00185-a, % !!!!!!!!!
       13-02-00419-a and 12-02-97006-r-povolzhye-a).
% and the Federal target program Research and
%Pedagogical Cadre for Innovative Russia (state contract
%no. 14.740.11.0800).
      The observations at the 6-m telescope are supported by the
      Ministry of Education and Science of Russian Federation (state contract
      no. 14.518.11.7070).
     Olga Maryeva thanks the grant of Dynasty Foundation.

\end{document}